\documentclass[12pt]{article}

\usepackage{graphicx}
\usepackage{epstopdf}
\usepackage{hyperref}

\def\onehalf{{\textstyle \frac12}}
\def\ii{{\rm i}}
\def\dd{{\rm d}}
\def\rr{{\bf r}}

\def\ssr#1{{\scriptscriptstyle\rm #1}}
\def\of#1{{{\scriptstyle(}#1{\scriptstyle)}}}
\def\oof#1{{{\scriptscriptstyle(}#1{\scriptscriptstyle)}}}
\def\abs#1{{{\scriptstyle|}#1{\scriptstyle|}}}
\def\aabs#1{{{\scriptscriptstyle|}#1{\scriptscriptstyle|}}}

\def\totder#1#2{\frac{\dd #1}{\dd #2}}

\def\tsty#1#2{{\textstyle\frac{#1}{#2}}}

\def\jour#1#2#3#4{{\it #1{}} {\bf #2}, #3 (#4)}
\def\lab#1{\label{eq:#1}}
\def\rf#1{(\ref{eq:#1})}
\def\Lie#1{\hbox{\sf #1}}

\newcommand{\be}{\begin{equation}}
\newcommand{\ee}{\end{equation}}
\newcommand{\bea}{\begin{eqnarray}}
\newcommand{\eea}{\end{eqnarray}}

\begin{document}

\begin{center}
{\Large\bf New separated polynomial solutions to the Zernike system
	on the unit disk and interbasis expansion}
\end{center}

\bigskip
       George S.\ Pogosyan,\footnote{Departamento de Matem\'aticas, 
  Centro Universitario de Ciencias Exactas e Ingenier\'ias, 
	Universidad de Guadalajara, M\'exico; Yerevan State University, 
	Yerevan, Armenia; and Joint Institute for Nuclear Research, 
	Dubna, Russian Federation.} 
	 Kurt Bernardo Wolf,\footnote{Instituto de Ciencias F\'isicas, 
	 Universidad Nacional Aut\'onoma de M\'exico, Cuernavaca.} 
	 and Alexander Yakhno\footnote{Departamento de 
	Matem\'aticas, Centro Universitario de Ciencias Exactas e 
	Ingenier\'ias, Universidad de Guadalajara, M\'exico.}

\vskip1cm



\vskip1cm


\begin{abstract}
      The differential equation proposed by Frits Zernike 
      to obtain a basis of polynomial orthogonal 
      solutions on the the unit disk to classify 
      wavefront aberrations in circular pupils, is 
      shown to have a set of new orthonormal 
      solution bases, involving Legendre and 
      Gegenbauer polynomials, in non-orthogonal 
      coordinates close to Cartesian ones.
      We find the overlaps between the original Zernike 
      basis and a representative of the new set, 
      which turn out to be Clebsch-Gordan coefficients. 
\end{abstract}

\vskip1cm

\section{Introduction: the Zernike system}  \label{sec:one}

In 1934 Frits Zernike published a paper which gave
rise to phase-contrast microscopy \cite{Zernike34}.
This paper presented a differential equation of second 
degree to provide an orthogonal basis of polynomial 
solutions on the unit disk to describe wavefront 
aberrations in circular pupils. This basis
was also obtained in Ref.\ \cite{Bhatia-Wolf} 
using the Schmidt orthogonalization process, 
as its authors noted that the reason to set up 
Zernike's differential equation had not been 
clearly justified. The two-dimensional differential 
equation in $\rr=(x,y)$ that Zernike solved is 
\be 
       \widehat Z\Psi\of\rr := \Big(\nabla^2  
               -(\rr\cdot\nabla)^2 
               -2\,\rr\cdot\nabla \Big)
                       \Psi\of\rr = -E\, \Psi\of\rr,
                       \lab{Zernikeq}
\ee
on the unit disk ${\cal D}:=\{\abs\rr\le1\}$ and 
$\Psi\of\rr\in{\cal L}^2({\cal D})$ (once two parameters
had been fixed by the condition of self-adjointness). 
The solutions found by Zernike are separable in polar 
coordinates $(r,\phi)$, with Jacobi polynomials of degrees $n_r$ 
in the radius $r:=\abs\rr$ times trigonometric functions 
$e^{\ii m\phi}$ in the angle $\phi$. The solutions are thus 
classified by $(n_r,m)$, which add up to non-negative 
integers $n=2n_r+\abs{m}$, providing the quantized 
eigenvalues $E_n=n(n+2)$ for the operator in \rf{Zernikeq}. 

The spectrum
$(n_r,m)$ or $(n,m)$ of the Zernike system is exactly that of 
the two-dimensional quantum harmonic oscillator. 
This evident analogy with the quantum oscillator spectrum 
has been misleading, however. Two-term raising and 
lowering operators do not exist; only three-term 
recurrence relations have been found 
\cite{Koornwinder,Kintner,Wunsche,Shakibaei,Ismail}.
Beyond the rotational symmetry that explains
the multiplets in $\{m\}$, no Lie algebra has been 
shown to explain the symmetry hidden in the equal
spacing of $n$ familiar from the oscillator model.

In Refs.\ \cite{Zernike-1,Zernike-2} we have interpreted
Zernike's equation \rf{Zernikeq} as defining a classical and a
quantum system with a non-standard `Hamiltonian' 
$-\onehalf\hat Z$. This turns out to be interesting because 
in the classical system the trajectories turn out
to be closed ellipses, and in the quantum system 
this Hamiltonian partakes in a cubic Higgs superintegrable 
algebra \cite{Higgs}. 

\begin{figure}[t]
\centering
\includegraphics[scale=0.60]{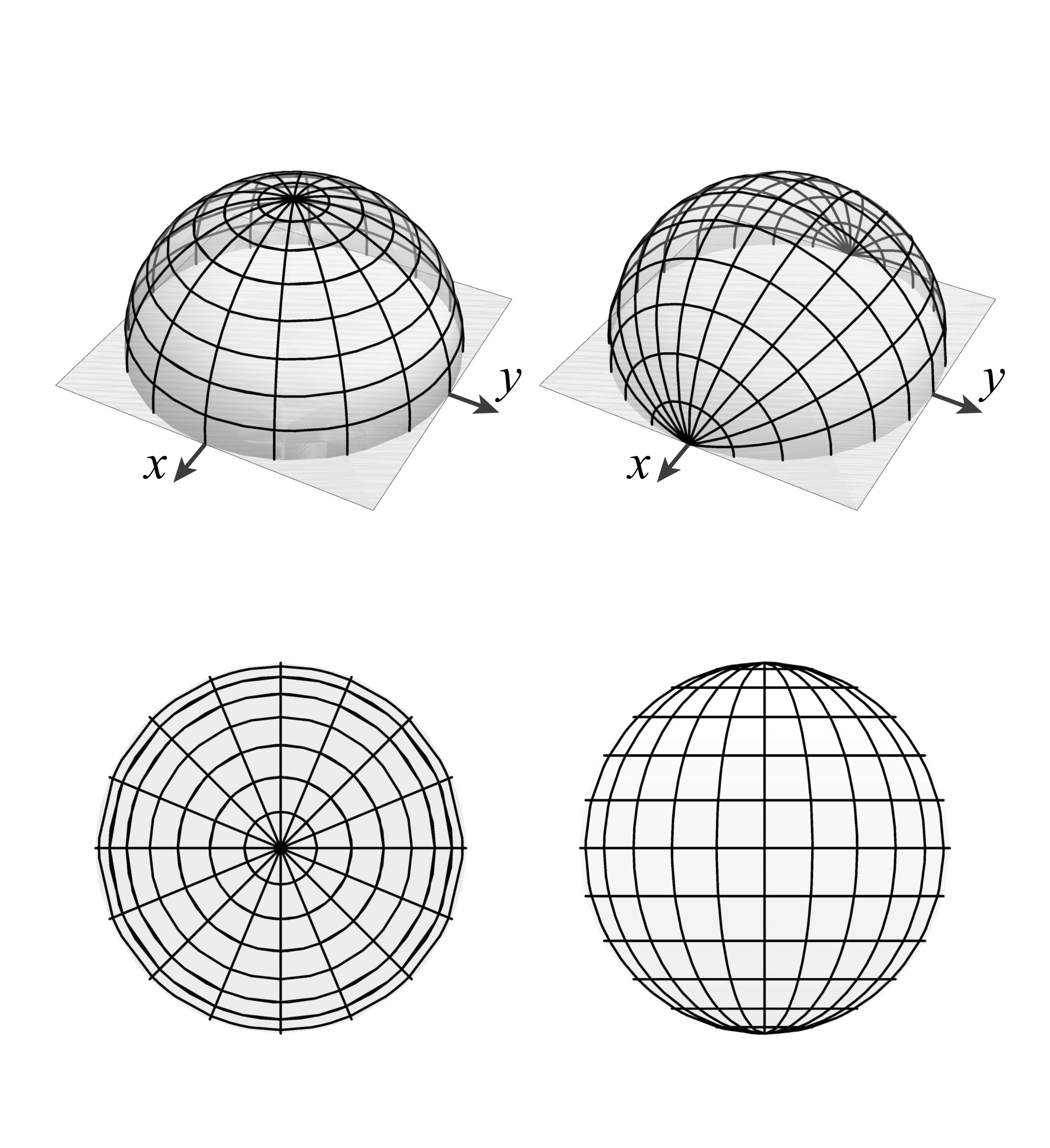}
\caption[]{{\it Top row\/}: Orthogonal 
coordinate systems that 
separate on the half-sphere ${\cal H}_+$. 
{\it Bottom row\/}: Their vertical 
projection on the disk $\cal D$. 
{\it Left\/}: Spherical 
coordinates with their pole at the 
$+z$-axis; separated solutions
will be marked by  I. 
{\it Right\/}: Spherical coordinates 
with their pole along the $+x$-axis, whose 
solutions are identified by II. 
The latter maps on non-orthogonal 
coordinates on the disk that also separate 
solutions of the Zernike equation.}
\label{fig:separ-coord}
\end{figure}

The key to solve the system was to perform a `vertical' 
map from the disk $\cal D$ in $\rr=(x,y)$
to a half-sphere in three-space $\vec{\,r}=(x,y,z)$, 
to be indicated as ${\cal H}_+:=\{\,|\!\vec{\,r}|=1,\,z\ge0\}$. 
On ${\cal H}_+$ the issue of separability of solutions 
becomes clear: the orthogonal spherical coordinate system 
$(\vartheta,\varphi)$, $\vartheta\in[0,\onehalf\pi]$, 
$\varphi\in(-\pi,\pi]$ on ${\cal H}_+$, projects
on the polar coordinates $(r,\phi)$ of $\cal D$.
But as shown in Fig.\ \ref{fig:separ-coord}, the 
half-sphere can also be covered with other orthogonal 
and {\it separated\/} coordinate systems (i.e., 
those whose boundary coincides with one fixed 
coordinate): where the coordinate poles are along the 
$x$-axis and the range of spherical angles is 
$\vartheta'\in[0,\pi]$ and $\varphi'\in[0,\pi]$. 
Since the poles of the spherical coordinates 
can lie in any direction of the 
$(x,y)$-plane and rotated around them, we
take the $x$-axis orientation as representing
the whole class of new solutions, which we
identify by the label II, to distinguish them
from Zernike's polar-separated solutions that
will be labelled I.

The coordinate system II is orthogonal on ${\cal H}_+$
but projects on {\it non\/}-ortho\-gonal 
ones on $\cal D$; the new separated solutions 
consist of Legendre and Gegenbauer polynomials 
\cite{Zernike-2}. Of course, the spectrum
$\{E_n\}$ in \rf{Zernikeq} is the same as in the
coordinate system I. Recall also that coordinates 
which separate a differential equation lead to 
extra commuting operators and constants of the motion. 
In this paper we proceed to find the I-II interbasis 
expansions between the original Zernike and the 
newly found solution bases; its compact expression 
in terms of \Lie{su($2$)} Clebsch-Gordan 
coefficients certainly indicates that some 
kind of deeper symmetry is at work.

The solutions of the Zernike system \cite{Zernike34} 
in the new coordinate system, that we indicate by  
$\Upsilon^\ssr{I}$ and $\Upsilon^\ssr{II}$ on 
${\cal H}_+$, and  $\Psi^\ssr{I}$ and $\Psi^\ssr{II}$ 
on ${\cal D}$, are succinctly derived and written 
out in Sect.\ \ref{sec:two}. In Sect.\ \ref{sec:three} 
we find the overlap between them, add some
remarks in the concluding Sect.\ \ref{sec:four},
and reserve for the Appendices some special-function
developments and an explicit list with the lowest-$n$ 
transformation matrices. 


\section{Two coordinate systems, two function bases} \label{sec:two}

The Zernike differential equation \rf{Zernikeq} in
$\rr=(x,y)$ on the disk ${\cal D}$ 
can be `elevated' to a differential equation on 
the half-sphere ${\cal H}_+$ in Fig.\ \ref{fig:separ-coord} 
through first defining the coordinates 
$\vec{\,\xi}=(\xi_1,\xi_2,\xi_3)$ by
\be 
	\xi_1:=x,\quad \xi_2:=y,\quad \xi_3:=\sqrt{1-x^2-y^2},
		\lab{xi-xyz}
\ee
then relating the measures of ${\cal H}_+$ and $\cal D$
through
\be 
	\dd^2S(\!\vec{\,\xi}) = \frac{\dd\xi_1\,\dd\xi_2}{\xi_3}
		= \frac{\dd x\,\dd y}{\sqrt{1-x^2-y^2}}
		= \frac{\dd^2\rr}{\sqrt{1-\abs{\rr}^2}},
			\lab{xi-xyz-measure}
\ee 
and the partial derivatives by 
$\partial_x=\partial_{\xi_1} - (\xi_1/\xi_3)\partial_{\xi_3}$
and $\partial_y=\partial_{\xi_2} - (\xi_2/\xi_3)\partial_{\xi_3}$.

\subsection{Map between ${\cal H}_+$ and $\cal D$ operators}

Due to the change in measure \rf{xi-xyz-measure}, the 
Zernike operator on $\cal D$, $\widehat Z$ in \rf{Zernikeq},
must be subject to a similarity transformation by 
the root of the factor between $\dd^2S(\vec\xi)$ and $\dd^2\rr$;
thus we define the Zernike operator on the half-sphere ${\cal H}_+$ and
its solutions, as  
\be 
	\widehat W:=(1-\abs\rr^2)^{1/4}\,\widehat Z\,(1-\abs\rr^2)^{-1/4},
		\quad \Upsilon(\!\vec{\,\xi}):=(1-\abs\rr^2)^{1/4}\Psi(\rr).
		\lab{Z-to-W}
\ee
In this way the inner product required for functions 
on the disk and on the sphere are related by
\be 
	(\Psi,\Psi')_{\cal D}:=\int_{\cal D}\dd^2\rr\,\Psi\of\rr^*\Psi'\of\rr
		= \int_{{\cal H}_+} \dd^2S(\!\vec{\,\xi})\,\Upsilon(\!\vec{\,\xi})^*\,\Upsilon'(\!\vec{\,\xi})
		=:(\Upsilon,\Upsilon')_{{\cal H}_+}.  \lab{Psi-Ups}
\ee

Perhaps rather surprisingly, the Zernike operator 
$\widehat W$ in \rf{Z-to-W} on 
$\vec{\,\xi}\in{\cal H}_+$  has the 
structure of ($-2$ times) a Schr\"odinger Hamiltonian,
\be 
	\widehat W \Upsilon(\!\vec{\,\xi})= \bigg(\Delta_\ssr{LB} 
	+ \frac{\xi_1^2+\xi_2^2}{4\xi_3^2}+1\bigg)\Upsilon(\!\vec{\,\xi})
	= -E\Upsilon(\!\vec{\,\xi}),
		\lab{hatW}
\ee
which is a sum of the Laplace-Beltrami operator
$\Delta_\ssr{LB}=\hat L^2_1+\hat L^2_2+\hat L^2_3$,
where 
\be 
	\hat L_1:=\xi_3\partial_{\xi_2}-\xi_2\partial_{\xi_3},\quad
	\hat L_2:=\xi_1\partial_{\xi_3}-\xi_3\partial_{\xi_1},\quad
	\hat L_3:=\xi_2\partial_{\xi_1}-\xi_1\partial_{\xi_1},
		\lab{LLL}
\ee
are the generators of a formal \Lie{so($3$)} Lie algebra.
The second summand in \rf{hatW} represents a radial potential
$V_\ssr{\!W}\of{r}:=-r^2/8(1-r^2)$ which has the form of a 
repulsive oscillator constrained to $(-1,1)$, whose rather 
delicate boundary conditions were addressed in Ref.\ \cite{Zernike-2}. 

The coordinates $\vec{\,\xi}$ can be now expressed in terms
of the two mutually orthogonal systems of coordinates 
on the sphere \cite{PSW1} as shown in Fig.\ \ref{fig:separ-coord}: 
\bea
       &&\hskip-30pt\hbox{System I:} \nonumber\\
               &&\hskip-30pt\xi_1 = \sin\vartheta\cos\varphi,\quad 
               \xi_2=\sin\vartheta\sin\varphi,\quad
               \xi_3=\cos\vartheta,\qquad
               \vartheta|_0^{\pi/2},\ \varphi|_{-\pi}^{\pi},
               \lab{Syst-I} \\
       &&\hskip-30pt\hbox{System II:} \nonumber\\
               &&\hskip-30pt\xi_1 = \cos\vartheta',\quad 
               \xi_2=\sin\vartheta'\cos\varphi',\quad
               \xi_3=\sin\vartheta'\sin\varphi',\qquad
               \vartheta'|_0^\pi,\ \varphi'|_0^\pi.
               \lab{Syst-II} 
\eea
In the following we succinctly give the normalized 
solutions for the differential equation \rf{hatW} 
in terms of the angles for ${\cal H}_+$ in the
coordinate systems I and II, and their projection 
as wavefronts on the disk $\cal D$ of the optical 
pupil. The spectrum of quantum numbers that 
classify each eigenbasis, $(n,m)$ and $(n_1,n_2)$, 
will indeed be formally identical with that of the 
two-dimensional quantum harmonic 
oscillator in polar and Cartesian coordinates,
respectively.


\subsection{Solutions in System I \rf{Syst-I}}

Zernike's differential equation \rf{Zernikeq} is
clearly invariant under rotations around the center
of the disk, corresponding to rotations of $\widehat W$
in \rf{hatW} around the $\xi_3$-axis of the coordinate system 
\rf{Syst-I} on the sphere. Written out in those
coordinates, it has the form of a Schr\"odinger equation, 
\be 
	\frac{1}{\sin\vartheta}\frac{\partial}{\partial \vartheta}
	\sin\vartheta\frac{\partial 
		\Upsilon^\ssr{I}(\vartheta,\varphi)}{\partial \vartheta}
	+\frac{1}{\sin^2\theta}\frac{\partial^2 
		\Upsilon^\ssr{I}(\vartheta,\varphi)}{\partial\varphi^2}
	+ (E + \tsty{1}{4}\tan^2\vartheta +1) 
	\Upsilon^\ssr{I}(\vartheta,\varphi) = 0,
		\lab{eq-I}
\ee
with a potential 
$V_\ssr{\!W}(\vartheta)=-\frac18\tan^2\vartheta$. 
Clearly this will separate into a differential equation 
in $\varphi$ with a separating constant $m^2$, where 
$m\in{\cal Z}:=\{0,\pm1,\pm2,\ldots\}$ and solutions 
$\sim e^{\ii m\varphi}$. This separation constant then
enters into a differential equation in $\vartheta$
that has also has the form of a one-dimensional 
Schr\"odinger equation with 
an effective potential of the P\"oschl-Teller type
$V^\ssr{I}_\ssr{\!eff}(\vartheta)
= (m^2-\frac14)\csc^2\vartheta - \frac14\sec^2\vartheta$,
whose solutions with the proper boundary conditions 
at $\vartheta=\onehalf\pi$ are Jacobi polynomials.

On the half-sphere the solutions to \rf{hatW} are thus
\be 
	\Upsilon^\ssr{I}_{n,m}(\vartheta,\varphi)
	:= \sqrt{\frac{n{+}1}{\pi}} 
	\, (\sin\vartheta)^{\abs{m}}\,(\cos\vartheta)^{1/2} 
			P_{\frac12(n-\aabs{m})}^{(\abs{m}, 0)}(\cos 2\vartheta)\,
			e^{\ii m \varphi}, \lab{YI}
\ee 
where $n\in{\cal Z}_0^+:=\{0,1,2,\ldots\}$ is the {\it principal\/}
quantum number corresponding to $E_n=n(n{+}2)$ in 
\rf{Zernikeq}. The index of the Jacobi polynomial is the 
{\it radial\/} quantum number that counts the number of radial
nodes, $n_r:= \frac12(n-\abs{m})\in{\cal Z}_0^+$. 
Thus, in each level $n$, the range of
angular momenta are $m\in\{-n,\,-n{+}2,\ldots,{n}\}$. 
These solutions are orthonormal over the half-sphere 
${\cal H}_+$ under the measure $\dd^2S^\ssr{I}(\vartheta,\varphi)
=\sin\vartheta\,\dd\vartheta\,\dd\varphi$ with the range 
of the angles $(\vartheta,\varphi)$ given in \rf{Syst-I}.

\begin{figure}[!h]
\centering
\includegraphics[scale=0.4]{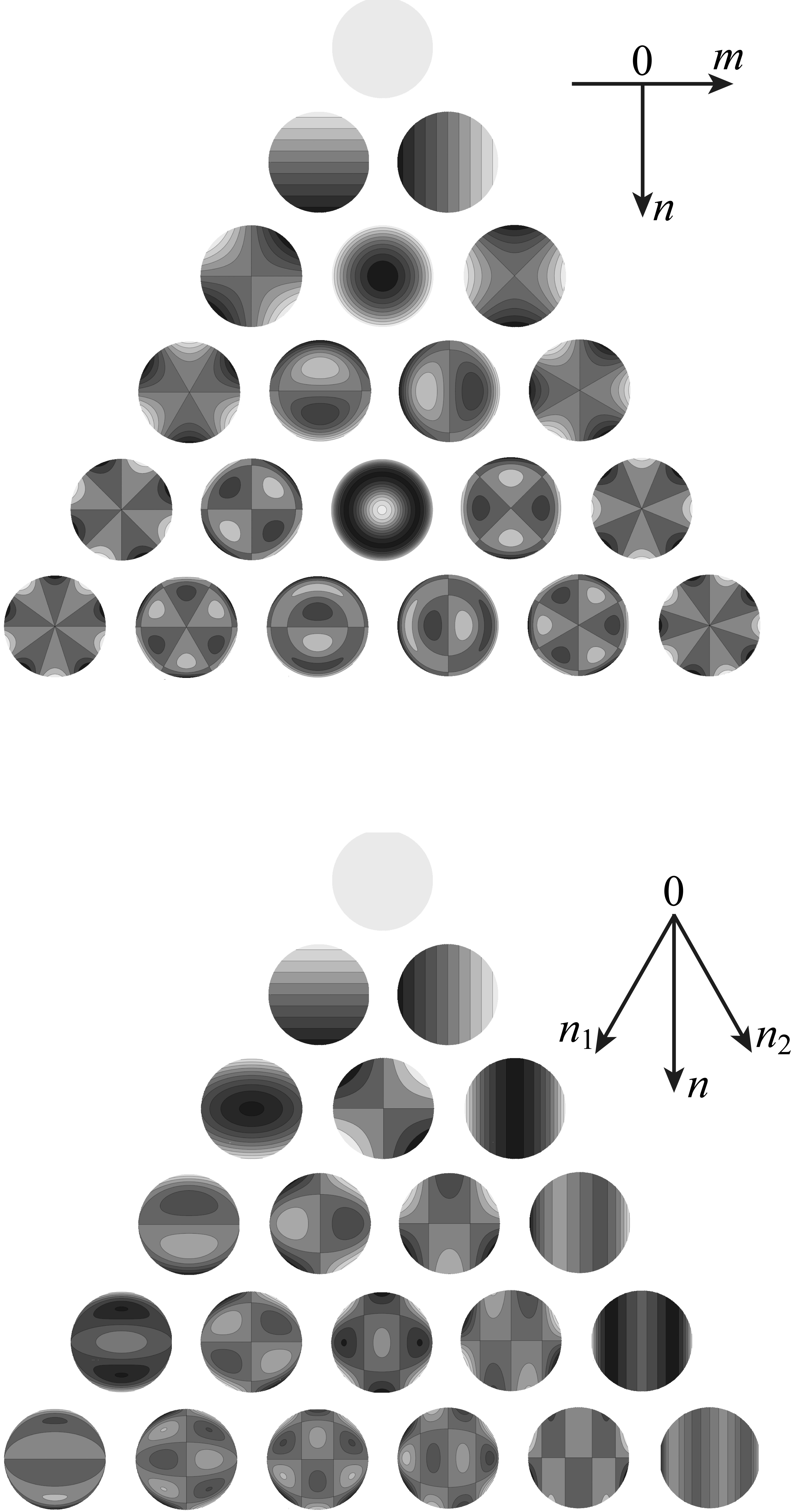}
\caption[]{{\it Top\/}: The basis of 
Zernike solutions $\Psi^\ssr{I}_{n,m}(r,\phi)$ in \rf{ourZernik}, 
normalized on the disk and classified by principal 
and angular momentum quantum numbers $(n,m)$. Since
they are complex, we show Re$\,\Psi^\ssr{I}_{n,m}$
for $m\ge0$ and Im$\,\Psi^\ssr{I}_{n,m}$ for $m<0$.
{\it Bottom\/}: The new real solutions $\Psi^\ssr{II}_{n,m}(r,\phi)$ 
in \rf{PsiII} of Zernike's equation \rf{Zernikeq} in 
the coordinate system II, classified by the quantum 
numbers $(n_1,n_2)$ (as if they were two-dimensional
quantum harmonic oscillator states ---which they are not).
Figure by Cristina Salto-Alegre.}
\label{fig:I-and-II}
\end{figure}

Projected on the disk $\cal D$ in polar coordinates $\rr=(r,\phi)$,
the original solutions of Zernike, orthonormal under 
the inner product in \rf{Psi-Ups}, are
\be 
	\Psi^\ssr{I}_{n,m}(r,\phi):= 	(-1)^{n_r}\sqrt{\frac{n+1}{\pi}}\,
	r^\aabs{m} P^{(\aabs{m},0)}_{n_r}(1{-}2r^2)\,
			e^{\ii m\phi},  \lab{ourZernik}
\ee
which are shown in Fig.\ \ref{fig:I-and-II} (top).

\subsection{Solutions in System II \rf{Syst-II}} 

The Zernike differential equation in the form \rf{hatW},
after replacement of the second coordinate system 
$(\vartheta',\varphi')$ in \rf{Syst-II} on the 
half-sphere, acting on functions separated as
\be 
	\Upsilon^\ssr{II}(\vartheta', \varphi') 
	= \frac{1}{\sqrt{\sin\vartheta'}}
		S(\vartheta')\,T(\varphi'),  \lab{eigenY}
\ee
yields a system of two simultaneous differential 
equations bound by a separation constant $k$, 
whose P\"oschl-Teller form is most evident in
variables $\mu=\onehalf\varphi'$ and 
$\nu=\onehalf\vartheta'$, 
\bea
	\totder{^2T(\mu)}{\mu^2}+\bigg(4k^2+\frac1{4\sin^2\mu}
		+\frac1{4\cos^2\mu}\bigg) T(\mu)&=&0, \lab{TT1}\\
	\totder{^2S(\nu)}{\nu^2}+\bigg( (E+1)^2
		+\frac{1-4k^2}{4\sin^2\nu}+\frac{1-4k^2}{4\cos^2\mu}\bigg) S(\nu)&=&0. 
		\lab{SS1} 
\eea
Finally, as shown in \cite{Zernike-1} and 
determined by the boundary conditions, 
two quantum numbers $n_1,n_2\in{\cal Z}_0^+$
are imposed for the solutions on ${\cal H}_+$,
yielding Gegenbauer polynomials in 
$\cos\vartheta'$ and Legendre polynomials in 
$\cos\varphi'$. These are  
\bea 
	\Upsilon^\ssr{II}_{n_1,n_2}(\vartheta', \varphi') 
		&:=&C_{n_1,n_2}\,(\sin\vartheta')^{n_1+1/2}
		(\sin\varphi')^{1/2}\, 
		C_{n_2}^{n_1+1}(\cos\vartheta')\,P_{n_1}(\cos\varphi'),
		\nonumber\\
			C_{n_1,n_2}&:=& 2^{n_1}n_1!
		\sqrt{\frac{(2n_1+1)(n_1+n_2+1)\,n_2!}{\pi\,(2n_1+n_2+1)!}},
			\lab{YII}
\eea
where the principal quantum number is 
$n=n_1+n_2\in{\cal Z}_0^+$, and with the 
$E=n(n{+}2)$ as before. 
The orthonormality of these solutions is
also over the half-sphere ${\cal H}_+$ 
under the formally same measure 
$\dd^2S^\ssr{II}(\vartheta',\varphi')
=\sin\vartheta'\,\dd\vartheta'\,\dd\varphi'$ 
where the angles have the range \rf{Syst-II}.

On the disk in Cartesian coordinates $\rr=(x,y)$, 
the solutions are 
\be 
	\Psi^\ssr{II}_{n_1,n_2}(x,y) =C_{n_1,n_2}\,(1-x^2)^{n_1/2}\,C^{n_1+1}_{n_2}(x)\,
			P_{n_1}\bigg(\frac{y}{\sqrt{1-x^2}}\bigg),
				\lab{PsiII}
\ee
separated in the non-orthogonal coordinates 
$x$ and $y/\surd(1-x^2)$, and normalized  under 
the inner product on $\cal D$ in \rf{Psi-Ups}.
These are shown in Fig.\ \ref{fig:I-and-II} (bottom).


\section{Expansion between I and II solutions}  \label{sec:three}

The two bases of solutions of the Zernike equation 
in the coordinate systems I and II on the half-sphere, 
$\Upsilon^\ssr{I}_{n,m}(\vartheta,\varphi)$ in \rf{YI} and
$\Upsilon^\ssr{II}_{n_1,n_2}(\vartheta',\varphi')$ in \rf{YII},
with the same principal quantum number $n$,
\be 
	\begin{array}{c}	n_1+n_2=n=2n_r+\abs{m}\in{\cal Z}_0^+,\\
		n_r,n_1,n_2\in{\cal Z}_0^+,\quad m\in\{-n,-n{+}2,\ldots,n\},
			\end{array} \lab{nnrmn1n2}
\ee
whose projections on the disk are shown in Fig.\ 
\ref{fig:I-and-II}, were arranged into pyramids with rungs 
labelled by $n$, and containing $n+1$ states each. 
They could be mistakenly seen as independent 
\Lie{su($2$)} multiplets of spin $j=\onehalf n$ because, 
as we said above, in system II they are not bases for this Lie algebra.
Nevertheless, in each rung $n$, the two bases must relate through linear
combination\footnote{The notation for the 
indices of the $\Upsilon^\ssr{I}$-function bases, here $(n,m)$, is 
different but equivalent to $(n_r,m)$ used in Ref.\ \cite{Zernike-2}.} 
\be 
	\Upsilon^\ssr{II}_{n_1,n_2}(\vartheta',\varphi') 
		=\!\!\! \sum_{m=-n\,\scriptscriptstyle(2)}^n \!\!\!W_{n_1,n_2}^{n,m}\,
		\Upsilon^\ssr{I}_{n,m}(\vartheta,\varphi), 	\lab{IItoI}
\ee
where $\sum_{m=-n\,\scriptscriptstyle(2)}^n$ indicates that $m$ takes
values separated by 2 as in \rf{nnrmn1n2}. 
The relation between the primed and unprimed angles in 
\rf{Syst-I} and \rf{Syst-II} is
\be 
	\begin{array}{rl}
	{}\quad\cos\vartheta'=\sin\vartheta\cos\varphi, &
	\sin\vartheta'=\sqrt{1-\sin^2\vartheta\cos^2\varphi},\\[3pt] \displaystyle
	\cos\varphi'= \frac{\sin\vartheta\sin\varphi}{\sqrt{1-\sin^2\vartheta\cos^2\varphi}},&
	\displaystyle \sin\varphi'= \frac{\cos\vartheta}{\sqrt{1-\sin^2\vartheta\cos^2\varphi}}.
		\end{array}  \lab{thphprthph}
\ee

To find the linear combination coefficients $W_{n_1,n_2}^{n,m}$
in \rf{IItoI}, we compute first the relation \rf{IItoI} near to the
boundary of the disk and sphere, at $\vartheta=\onehalf\pi-\varepsilon$
for small $\varepsilon$, so that 
$\cos\vartheta=-\sin\varepsilon\approx-\varepsilon$ and
$\sin\vartheta=\cos\varepsilon\approx1-\onehalf\varepsilon^2$.
There, \rf{thphprthph} becomes
\be 
	\begin{array}{rl}
	{}\quad\cos\vartheta'\approx\cos\varphi, &
	\sin\vartheta'\approx\sin\varphi,\\[3pt] \displaystyle
	\cos\varphi'\approx\cos\varepsilon,&
	\displaystyle \sin\varphi'\approx -\sin\varepsilon/\sin\varphi.
		\end{array}  \lab{limthphprthph}
\ee
Hence, when $\varepsilon\to0$ is at the rim of the disk and sphere,
after dividing \rf{IItoI} by $\surd(-\varepsilon)$ on both sides, 
this relation reads
\be 
	C_{n_1,n_2}\,(\sin\varphi)^{n_1} C^{n_1+1}_{n_2}(\cos\varphi)\,P_n(1)
	= \sqrt{\frac{n{+}1}{\pi}}\!\!\! \sum_{m=-n\,\scriptscriptstyle(2)}^n 
			\!\!\!W_{n_1,n_2}^{n,m}\,
		 P^{(\aabs{m},0)}_{n_r}(-1)\,e^{\ii m\varphi},
		\lab{Upsateps0}
\ee
with $n_r=\frac12(n-\aabs{m})$. Recalling that 
$P_{n_1}(1)=1$ and $P^{(\aabs{m},0)}_{n_r}(-1)=(-1)^{n_r}$, 
we can now use the orthogonality of the $e^{\ii m\varphi}$ 
functions to express the interbasis coefficients as a Fourier
integral,
\be 
	W_{n_1,n_2}^{n,m}= \frac{(-1)^{n_r}\,C_{n_1,n_2}}{2\sqrt{\pi(n{+}1)}}
		\int_{-\pi}^\pi \dd\varphi\, (\sin\varphi)^{n_1}
			C^{n_1+1}_{n_2}(\cos\varphi)\,\exp(-\ii m\varphi).
				\lab{Wintexp}
\ee

The integral \rf{Wintexp} does not appear as such
in the standard tables \cite{GR}; in Appendix A we
derive the result and show that it can be written
in terms of a hypergeometric ${}_3F_2$ polynomial 
which are \Lie{su($2$)} Clebsch-Gordan coefficients
of a special structure,
\bea
	&&W^{n, m }_{n_1, n_2}  
       =  \frac{\ii^{n_1} (-1)^{(m+|m|)/{2}}\,n_1!\,(n_1+n_2)!
       	}{\Big(\onehalf(n_1{+}n_2{+}m)\Big)!\,\Big(\onehalf({n_1{-}n_2{-}m})\Big)!}
	\, \sqrt{\frac{2n_1+1}{n_2!\,(2n_1+ n_2+1)!}}
			\nonumber\\[3mm]
	&&{\qquad\qquad\qquad}\times {_3F_2}\left(
		{ -n_2, \quad n_1 + 1, \quad -\onehalf(n_1+n_2+m) 
			\atop -n_1-n_2,\quad\onehalf(n_1-n_2 - m)+ 1} \Bigg|\,\,1\right)
				\lab{coef-01}\\[5pt]
	&&\phantom{W^{n, m }_{n_1, n_2}\,}=\ii^{n_1} (-1)^{(m+|m|)/{2}}\, 
		C^{n_1,\ 0}_{\frac12 n,\, -\frac12 m;\ \frac12 n,\, \frac12 m}, \lab{CGC-01}
\eea
where we have used the notation of
Varshalovich {\it et al.}\ in Ref.\ \cite{Varshalovich}
that couples the \Lie{su($2$)} states $(j_1,m_1)$ and 
$(j_2,m_2)$ to $(j,m)$, as $C^{j,m}_{j_1,m_1;\,j_2,m_2} 
	\equiv C{\textstyle{j_1,\atop m_1,}{j_2,\atop m_2,}{j\atop m}}
	\equiv \langle j_1,m_1;\,j_2,m_2|(j_1,j_2)\,j,m\rangle. $ 

One can then use the orthonormality properties of the 
Clebsch-Gordan coefficients to  write
the transformation inverse to \rf{IItoI} as
\bea 
	\Upsilon^\ssr{I}_{n,m}(\vartheta,\varphi) 
		&=& \sum_{n_1=0}^n  \widetilde{W}^{n_1,n_2}_{n,m}\,
		\Upsilon^\ssr{II}_{n_1,n_2}(\vartheta',\varphi'),
			\lab{ItoII} \\
	\widetilde{W}_{n, m }^{n_1, n_2}&=&(-\ii)^{n_1} (-1)^{(m+|m|)/{2}}\, 
		C^{n_1,\ 0}_{\frac12 n,\, -\frac12 m;\ \frac12 n,\, \frac12 m},	
\eea
with $n_1+n_2=n$. The relation between the unprimed and primed
angles of the coordinate systems I and II is the inverse of 
\rf{thphprthph}, namely
\be 
	\begin{array}{rl}
	{}\quad\cos\vartheta=\sin\vartheta'\sin\varphi', &
	\sin\vartheta=\sqrt{1-\sin^2\vartheta'\sin^2\varphi'},\\[3pt] \displaystyle
	\cos\varphi= \frac{\cos\vartheta'}{\sqrt{1-\sin^2\vartheta'\sin^2\varphi'}},&
	\displaystyle \sin\varphi= \frac{\sin\vartheta'}{\sqrt{1-\sin^2\vartheta'\sin^2\varphi'}}.
		\end{array}  \lab{invththp}
\ee


\section{Concluding remarks}   \label{sec:four}

The new polynomial solutions of the Zernike 
differential equation \rf{Zernikeq} can be of 
further use in the treatment of generally 
off-axis wavefront aberrations in circular pupils. 
While the original basis of Zernike polynomials
$\Psi^\ssr{I}_{n,m}(r,\phi)$ serves naturally
for axis-centered aberrations, the new basis
$\Psi^\ssr{II}_{n_1,n_2}(r,\phi)$ in \rf{PsiII} 
includes, for $n_1=0$, plane wave-trains with 
$n_2$ nodes along the $x$-axis of the pupil, 
which are proportional to  $U_{n_2}(x)$, the 
Chebyshev polynomials of the second kind.

We find that the Zernike system is also very 
relevant for studies of `non-standard' 
symmetries described by Higgs algebras.
While rotations in the basis of spherical
harmonics is determined through the 
Wigner-$D$ functions \cite{Varshalovich} of
the rotation angles on the sphere, here the
boundary conditions of the disk and sphere
allow for only a $\onehalf\pi$-rotation of
the $z$-axis to orientations in $x$--$y$ plane,
and the basis functions do not relate through
Wigner $D$-functions, but Clebsch-Gordan
coefficients of a special type. Since the
classical and quantum Zernike systems analysed
in \cite{Zernike-1,Zernike-2} have several new and
exceptional properties, we surmise that 
applications not yet evident in this paper must
also be of interest.


\section*{Appendix A. The integral \rf{Wintexp} and Clebsch-Gordan coefficients}
	The integral in \rf{Wintexp} does not
seem to be in the literature, although similar integrals appear
in a paper of Kildyushov \cite{Kildyushov} to calculate his
{\it three\/} coefficients. Thus let us solve {\it ab initio}, with 
$\lambda=n_1$ and $\nu=n_2$, integrals of the kind
\be 
	I_\nu^{\lambda,m}:=\int_{-\pi}^{\pi} \dd\varphi\,
		\sin^{\lambda}\! \varphi\,\, 
		C_{\nu}^{\lambda+1} (\cos\varphi) \, e^{-\ii m \varphi},
	\quad \lambda,\nu \in\{0,1,2,\ldots\}. \lab{Ilnm}
\ee

We write the trigonometric function and the Gegenbauer polynomial 
in their Fourier series expansions,
\bea
	\sin^{\lambda} \varphi 
	&=& \frac{e^{\ii \lambda \varphi}}{(2\ii)^{\lambda}} (1-e^{-2\ii\varphi })^{\lambda}
	=\frac1{(2\ii)^\lambda}  \sum\limits_{k=0}^{\lambda} 
		\frac{(-1)^k\, \lambda!}{k! \, (\lambda-k)! }\, e^{\ii(\lambda-2 k)\varphi},
				\lab{APP-2-01}\\[3mm]
	C^{\lambda+1}_{\nu}(\cos\varphi) 
	&=& \sum\limits_{l=0}^{\nu} \frac{(\lambda +l)!}{l! \, (\nu - l)!} 
	\frac{(\lambda + \nu - l)!}{(\lambda !)^2} e^{-\ii(\nu-2l)\varphi}.
				\lab{APP-3-01}
\eea
Substituting these expansions in \rf{Ilnm}, 
using the orthogonality of the 
$e^{\ii\kappa\varphi}$ functions and thereby
eliminating one of the two sums, we find a ${}_3F_2$
hypergeometric series for unit argument,
\bea
	&&I_\nu^{\lambda,m}= \frac{2\pi}{(2\ii)^{\lambda}} \frac{(\lambda{+}\nu)!}{\nu!} 
		\frac{(-1)^{(\lambda-\nu - m)/2}}{\Big(\onehalf(\lambda-\nu - m)\Big)!\, 
			\Big( \onehalf(\lambda+\nu + m)\Big)!} \nonumber\\[3mm]
	&&{\qquad\qquad}\times {_3F_2} 
	\left({-\nu, \quad \lambda+1, \quad -\onehalf(\lambda+\nu+m) 
		\atop -\lambda-\nu, \quad  \onehalf(\lambda -\nu - m) + 1}\Bigg|\,\,1\right).
				\lab{APP-4-01}
\eea
Multiplying this by the coefficients $C_{n_1,n_2}$ in  
\rf{Wintexp}, one finds the first expression in \rf{coef-01}.

In order to relate the previous result with the 
\Lie{su($2$)} Clebsch-Gordan coefficients in 
\rf{CGC-01}, we use the formula in 
\cite[Eq.\ (21), Sect.\ 8.2]{Varshalovich} 
for the particular case at hand, and a 
relation between $_3F_2$-hypergeometric functions,
\be 
	{_3F_2}\left({ a,\ b,\ c	\atop d,\ e} \Bigg|\,\,1\right)
	=\frac{\Gamma(d)\,\Gamma(d{-}a{-}b)}{\Gamma(d{-}a)\,\Gamma(d{-}b)}\,\,
	{_3F_2}\left({ a,\ b,\ e{-}c	\atop a{+}b{-}d{+}1,\ e} \Bigg|\,\,1\right),
		\lab{F32}
\ee
to write these particularly symmetric coefficients as
\bea 
	&&C^{\gamma,0}_{\alpha,\,-\beta;\ \alpha,\,\beta}
	= \frac{(2\alpha)!\,\gamma!}{(\alpha+\beta)!\,(\gamma-\alpha-\beta)!}
	\sqrt{\frac{2\gamma+1}{(2\alpha-\gamma)!\,(2\alpha+\gamma+1)!}}\nonumber\\
	&&{\qquad\qquad\qquad}\times {_3F_2}\left(
		{ -2\alpha+\gamma,\quad \gamma+1,\quad -\alpha-\beta
			\atop -2\alpha,\quad \gamma-\alpha-\beta+1} \Bigg|\,\,1\right).
				\lab{Cabc}	
\eea
Finally, upon replacement of $\gamma=n_1$, 
$\alpha=\onehalf n=\onehalf(n_1+n_2)$ and
$\beta=\onehalf m$, the expression 
\rf{coef-01} reduces to \rf{CGC-01}
times the phase and sign.


\section*{Appendix B. The lowest $W^{n,m}_{n_1,n_2}$ coefficients}

The interbasis expansion coefficients binding the 
two bases in \rf{IItoI} and Fig.\ \ref{fig:I-and-II},
can be seen as $(n+1)\times(n+1)$ matrices 
${\bf W}_{\!\!\oof{n}}=\Vert W^{n,m}_{n_1,n_2}\Vert$ with composite 
rows $(n_1,n_2)$ and columns $(n,m)$
for each rung $n_1+n_2=n\in{\cal Z}_0^+$, on
$(n+1)$-dimensional column vectors of functions  
as ${\bf\Upsilon}^\ssr{II}(\vartheta',\varphi')
={\bf W}_{\!\!\oof{n}}{\bf\Upsilon}^\ssr{I}(\vartheta,\varphi)$. 
The elements $W^{n,m}_{n_1,n_2}$ in \rf{CGC-01}
are the product of phases
\be 
	\ii^{n_1},\qquad 
	(-1)^{\pm\frac12(m+\aabs{m})}
	=\left\{ \begin{array}{ll}(-1)^m,& m>0,\\ 1,& m\le0, \end{array}\right.
		\lab{signss}
\ee
times the special Clebsch-Gordan coefficients $C^{n_1,0}_{\frac12 n,-\frac12 m;\frac12 n,\frac12 m}$

For the first five rungs in Fig.\ \ref{fig:I-and-II}, these are

\medskip

\noindent For $n=0$:\quad $W_{\!\!(0)}= C^{0,0}_{0,0;0,0} = 1$,
\bea
	&&{\hskip-15pt}{\bf W}_{\!\!(1)}=
		 \bordermatrix{&m=1&-1\cr
		n_1{=}1& -\ii C^{1,0}_{\frac12,-\frac12;\frac12,\frac12}&\ii C^{1,0}_{\frac12,\frac12;\frac12,-\frac12}\cr
		{}\hfill 0& -C^{0,0}_{\frac12,-\frac12;\frac12,\frac12}& C^{0,0}_{\frac12,\frac12;\frac12,-\frac12}\cr},
			\nonumber\\ 
	&&{\hskip-20pt}{\bf W}_{\!\!(2)}=		
		 \bordermatrix{&m=2&0&-2\cr 		 
	n_1{=}2&-C^{2,0}_{1,-1;1,1}&-C^{2,0}_{1,0;1,0}&-C^{2,0}_{1,1;1,-1}\cr
	{}\hfill 1&\ii C^{1,0}_{1,-1;1,1}&      0     &\ii C^{1,0}_{1,1;1,-1}\cr
	{}\hfill 0&C^{0,0}_{1,-1;1,1}&C^{0,0}_{1,0;1,0}&C^{0,0}_{1,1;1,-1}\cr},
			\nonumber\\
	&&{\hskip-20pt}{\bf W}_{\!\!(3)}={}\hskip-0.5cm
		 \bordermatrix{&m=3&1&-1&-3\cr 
	n_1{=}3&\ii C^{3,0}_{\frac32,-\frac32;\frac32,\frac32}&\ii C^{3,0}_{\frac32,-\frac12;\frac32,\frac12}
		   &-\ii C^{3,0}_{\frac32,\frac12;\frac32,-\frac12}&-\ii C^{3,0}_{\frac32,\frac32;\frac32,-\frac32}\cr	
	{}\hfill 2&C^{2,0}_{\frac32,-\frac32;\frac32,\frac32}&C^{2,0}_{\frac32,-\frac12;\frac32,\frac12}
		   &-C^{2,0}_{\frac32,\frac12;\frac32,-\frac12}&-C^{2,0}_{\frac32,\frac32;\frac32,-\frac32}\cr		
	{}\hfill 1&-\ii C^{1,0}_{\frac32,-\frac32;\frac32,\frac32}&-\ii C^{1,0}_{\frac32,-\frac12;\frac32,\frac12}
		   &\ii C^{1,0}_{\frac32,\frac12;\frac32,-\frac12}&\ii C^{1,0}_{\frac32,\frac32;\frac32,-\frac32}\cr		
	{}\hfill 0&-C^{0,0}_{\frac32,-\frac32;\frac32,\frac32}&-C^{0,0}_{\frac32,-\frac12;\frac32,\frac12}
		   &C^{0,0}_{\frac32,\frac12;\frac32,-\frac12}&C^{0,0}_{\frac32,\frac32;\frac32,-\frac32}\cr},
		   	\nonumber\\						
	&&{\hskip-20pt}{\bf W}_{\!\!(4)}={}\hskip-0.5cm
		 \bordermatrix{&m=4&2&0&-2&-4\cr 
	n_1{=}4&C^{4,0}_{2,-2;2,2}&C^{4,0}_{2,-1;2,1}&C^{4,0}_{2,0;2,0}&C^{4,0}_{2,1;2,-1}&C^{4,0}_{2,2;2,-2}\cr 
	{}\hfill 3&-\ii C^{3,0}_{2,-2;2,2}&-\ii C^{3,0}_{2,-1;2,1}&0&-\ii C^{3,0}_{2,1;2,-1}&-\ii C^{3,0}_{2,2;2,-2}\cr{}\hfill 
	2&-C^{2,0}_{2,-2;2,2}&-C^{2,0}_{2,-1;2,1}&-C^{2,0}_{2,0;2,0}&-C^{2,0}_{2,1;2,-1}&-C^{2,0}_{2,2;2,-2}\cr 
	{}\hfill  1&\ii C^{1,0}_{2,-2;2,2}&\ii C^{1,0}_{2,-1;2,1}&0&\ii C^{1,0}_{2,1;2,-1}&\ii C^{1,0}_{2,2;2,-2}\cr {}\hfill 
	0&C^{0,0}_{2,-2;2,2}&C^{0,0}_{2,-1;2,1}&C^{0,0}_{2,0;2,0}&C^{0,0}_{2,1;2,-1}&C^{0,0}_{2,2;2,-2}\cr}, 
		\nonumber
\eea
where some elements are zero because 
$C^{n_1,0}_{\frac12n,0;\,\frac12n,0}=0$ 
for even $n$ and odd $n_1$.

The Zernike polynomials come in complex conjugate pairs, 
$\Upsilon^\ssr{I}_{n,m}=\Upsilon^\ssr{I\,*}_{n,-m}$,
while the $\Upsilon^\ssr{II}_{n_1,n_2}$'s are real.
The linear combinations afforded by the ${\bf W}$ matrices
above indeed yield real functions because 
\be 
	 C^{n_1,0}_{\frac12n,-\frac12m;\,\frac12n,\frac12m}
		=(-1)^{n_2} C^{n_1,0}_{\frac12n,\frac12m;\,\frac12n,-\frac12m}.
				  \lab{sym2}
\ee

\section*{Acknowledgements}

We thank Prof.\ Natig M.\ Atakishiyev for his interest in
the matter of interbasis expansions, and acknowledge the 
technical help from Guillermo Kr\"otzsch ({\sc icf-unam}) and
Cristina Salto-Alegre with the figures. 
G.S.P.\ and A.Y.\ thank the support of project 
{\sc pro-sni-2017} (Universidad de Guadalajara). 
N.M.A. and K.B.W.\ acknowledge the support of {\sc unam-dgapa} 
Project {\it \'Optica Matem\'atica\/} {\sc papiit}-IN101115.


\end{document}